\documentclass[twocolumn,amsmath,amssymb,showpacs,prb]{revtex4}

\usepackage{graphicx} 	
\usepackage{bm} 		
\usepackage{amssymb}
\usepackage{amsmath}

\begin{document}


\title{Self-consistent solution for proximity effect and Josephson current in ballistic graphene SNS Josephson junctions}

\author{Annica M. Black-Schaffer}
 \affiliation{Department of Applied Physics, Stanford University, Stanford, California 94305}
 \author{Sebastian Doniach}
 \affiliation{Departments of Physics and Applied Physics, Stanford University, Stanford, California 94305}

\date{\today}
       
\begin{abstract}
We use a tight-binding Bogoliubov-de Gennes (BdG) formalism to self-consistently calculate the proximity effect, Josephson current, and local density of states in ballistic graphene SNS Josephson junctions. Both short and long junctions, with respect to the superconducting coherence length, are considered, as well as different doping levels of the graphene. We show that self-consistency does not notably change the current-phase relationship derived earlier for short junctions using the non-selfconsistent Dirac-BdG formalism\cite{Titov06} but predict a significantly increased critical current with a stronger junction length dependence. In addition, we show that in junctions with no Fermi level mismatch between the N and S regions superconductivity persists even in the longest junctions we can investigate, indicating a diverging Ginzburg-Landau superconducting coherence length in the normal region.
\end{abstract}


\maketitle
%
%
\section{Introduction}
Recently SNS graphene Josephson junctions have been experimentally realized by depositing $s$-wave superconducting contacts on top of a graphene sheet.\cite{Heersche07,Shailos06, Du07} These junctions have been shown to carry a Josephson current that, depending on the position of the Fermi level relative to the Dirac point, consists of either electrons or holes, and, even at the Dirac point, where the density of states is zero, a finite supercurrent has been measured. 

Theoretically, a finite Josephson current was predicted by Titov {\it et.~al} \cite{Titov06} shortly before the first experimental realization. 
This result was based on the Dirac-Bogoliubov-de Gennes (DBdG) formalism developed by Beenakker\cite{Beenakker06condmat} where the band structure is approximated by the particular double Dirac cone spectra found in graphene and the traditional BdG formalism is then applied to solve for the wave functions throughout the junction.
The Josephson current was approximated as the current carried by the subgap bound Andreev states in the junction, an approximation valid in the short junction regime\cite{Beenakker91} where the length $L$ of the junction is smaller than the superconducting coherence length $\xi$.  Graphene SBS junctions, where B is a barrier created by a heavily doped graphene slip, were later also investigated using the same general formalism.\cite{Maiti06}

The DBdG approach in Ref.~\onlinecite{Titov06} neglects the spacial dependence of the superconducting order parameter as it assumes a constant, non-zero, order parameter in the superconductor and an abrupt change to zero at the SN interface. 
However, close to a SN interface the superconducting order parameter is expected to vary strongly as a function of the distance to the interface.
In this paper we address this problem by self-consistently calculating the superconducting order parameter in graphene SNS Josephson junctions using the tight-binding Bogoliubov-de Gennes (TB BdG) formalism.
Specifically, this allows us to explicitly calculate the proximity effect depletion of the order parameter close to the interface and it also results in a Josephson current properly calculated from the superconducting proximity effect. In addition, the results will no longer be limited to the short junction regime. 

\begin{figure}[htb]
\includegraphics[scale = 1]{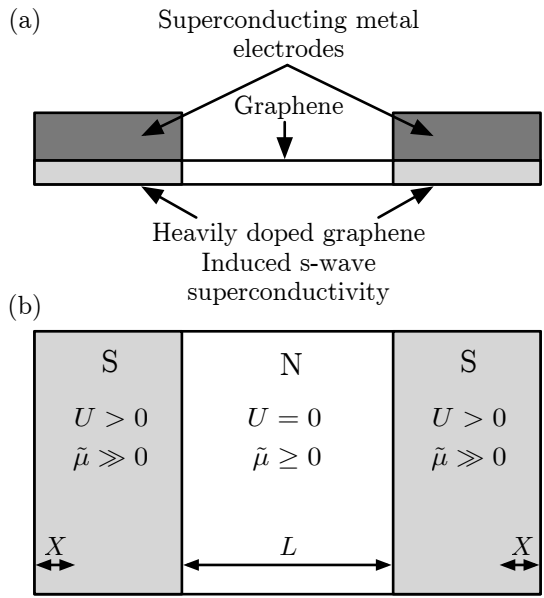}
\caption{\label{fig:SNSjunction} Schematic of (a) an experimental SNS graphene Josephson junction and (b) the model setup with input parameters: pairing potential $U$, effective chemical potential $\tilde{\mu}$, length of normal region $L$, and region where the phase of the order parameter will be kept fixed $X$.}
\end{figure}

As in the DBdG approach, we will study an impurity free, ballistic, graphene sheet by itself, and therefore need to model the influence of the superconducting contacts on the graphene. We do this by assuming the following two effects on the graphene directly under the superconducting contacts: The superconducting contacts induce 1) an on-site attractive Hubbard pairing potential $U$, which will lead to a $s$-wave superconducting order parameter $\Delta_U$, and 2) a heavy doping in the graphene.
The first assumption implies that the superconducting contacts induce a pairing potential in the graphene but that the superconducting order parameter itself in the graphene junction will be subjected to the proximity effect when solved for self-consistently.
Since solving self-consistently not only for the order parameter but also for the chemical potential in the system adds a significant amount of complexity to the problem, we will assume that the doping profile throughout the junction is set by different, but constant, effective chemical potentials $\tilde{\mu}$ in the S and N regions. The same assumption was used in the non-self-consistent DBdG approach.
Figure \ref{fig:SNSjunction} shows a schematic of the experimental (a) and model setup (b). The interface between S and N is assumed to be clean and smooth, a realistic assumption as experiments has indicated a high transparency of the SN interfaces\cite{Heersche07,Du07}.
%
%
%

Using our self-consistent TB BdG formalism we find that the functional dependence of the supercurrent on the phase drop across the junction is not significantly altered from the non-self-consistent DBdG results by Titov {\it et.~al} \cite{Titov06} and in fact extends to junctions with $L > \xi$. 
However, we find that the absolute value of the critical current is in general significantly higher and has a stronger $L$-dependence, especially for junctions where the N region is moderately doped.
In addition, we show that for junctions with no Fermi level mismatch (FLM) between the N and S regions superconductivity persists, not just proximity-wise, even in the longest junctions we can study. 
Finally, we also report detailed local density of states (LDOS) plots for different SN junctions, showing the evolution of the superconducting energy gap throughout the junction. These results should directly be comparable with experimental data obtained using a point contact scanning tunneling probe. 
%

Before proceeding it is worth noting that the TB BdG formalism not only effectively captures the proximity effect in a $s$-wave graphene Josephson junction but is also easily extendable to include other short-range electronic coupling terms in the graphene. 
Going back originally to Linus Pauling\cite{Paulingbook} $p\pi$-bonded planar organic molecules, such as graphene, have been recognized to have enhanced nearest-neighbor spin-singlet bonds compared to polar configurations. In a Hamiltonian formulation this spin-singlet enhancement takes the form of an intrinsic $J {\bf S}_i\cdot {\bf S}_j$ term between nearest neighbors which for strong enough coupling will give rise to mean-field superconductivity. In earlier work\cite{Black-Schaffer07} we have shown that in the bulk this gives rise to a time-reversal symmetry breaking $d_{x^2-y^2}+id_{xy}$ superconducting order parameter.  In addition, a $J {\bf S}_i\cdot {\bf S}_j$ term can also be used to model $d$-wave superconducting contacts in graphene Josephson junctions.
Self-consistent studies including these intrinsic spin-singlet electronic correlations in the graphene inside a $s$-wave SNS Josephson junction as well as the effect of $d$-wave contacts are the subject of future publications.
%
\section{Method}
\begin{figure}[htb]
\includegraphics[scale = 1]{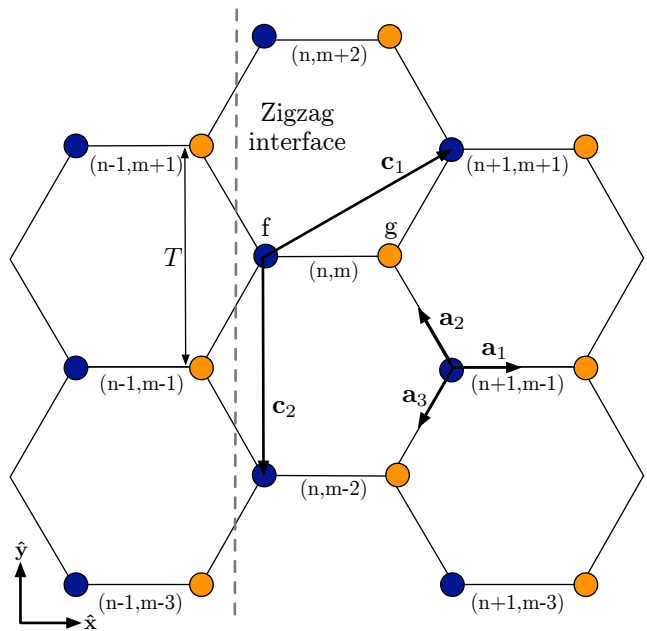}
\caption{\label{fig:graphene} (Color online) The graphene honeycomb lattice with the two different atomic sites $f$ and $g$, the unit cell vectors $\{{\bf c}_1,{\bf c}_2\}$, the three nearest neighbor directions $\{{\bf a}_1,{\bf a}_2,{\bf a}_3\}$, the zigzag interface with its $T = a = 2.46$~\AA~long unit cell in the $\hat{\bf y}$-direction, and the $(n,m)$ notation for labeling each unit cell.}
\end{figure}

Based on the motivation above we model a graphene SNS Josephson junction using the following effective tight-binding, attractive Hubbard Hamiltonian
%
\begin{align}
\label{eq:H_eff}
H_{\rm eff} = & -t \!\! \! \! \sum_{<{\bf i},{\bf j}>,\sigma} \!\!\! (f_{{\bf i}\sigma}^\dagger g_{{\bf j}\sigma} + g_{{\bf i}\sigma}^\dagger f_{{\bf j}\sigma}) +
\sum_{{\bf i},\sigma} \tilde{\mu}({\bf i})(f_{{\bf i}\sigma}^\dagger f_{{\bf i}\sigma} + g_{{\bf i}\sigma}^\dagger g_{{\bf i}\sigma}) \nonumber \\
& -\sum_{\bf i} U({\bf i}) (f_{{\bf i}\uparrow}^\dagger f_{{\bf i}\uparrow}f_{{\bf i}\downarrow}^\dagger f_{{\bf i}\downarrow} + g_{{\bf i}\uparrow}^\dagger g_{{\bf i}\uparrow}g_{{\bf i}\downarrow}^\dagger g_{{\bf i}\downarrow}).
\end{align}
Here $f_{{\bf i}\sigma}^\dagger$ is the creation operator on the A-site in cell ${\bf i} = (n,m)$ of the honeycomb lattice, and $g_{{\bf i}\sigma}^\dagger$ on the B-site of the same unit cell ${\bf i}$, see Fig.~\ref{fig:graphene}. $<\! {\bf i},{\bf j} \!>$ indicates a sum over nearest neighbors. The energy parameters are the hopping energy $t \approx 2.5$~eV which we will assume constant throughout the junction, the on-site attractive Hubbard term $U({\bf i})$ which is only non-zero in the S regions, and the effective chemical potential $\tilde{\mu}({\bf i})$ which includes the combined effect of the chemical potential $\mu$, which is constant in the whole system, and the local doping. 
%
To proceed, we use the Hartree-Fock-Bogoliubov mean-field approximation to arrive at the following one-particle Hamiltonian
%
\begin{align}
\label{eq:H_mf}
H_{{\rm MF}} = & -t \!\! \! \!  \sum_{<{\bf i},{\bf j}>,\sigma} \!\! \! (f_{{\bf i}\sigma}^\dagger g_{{\bf j},\sigma} + g_{{\bf i}\sigma}^\dagger f_{{\bf j},\sigma}\!) +
\! \sum_{{\bf i},\sigma} \tilde{\mu}({\bf i})(f_{{\bf i}\sigma}^\dagger f_{{\bf i}\sigma} + g_{{\bf i}\sigma}^\dagger g_{{\bf i}\sigma}\!) \nonumber \\
& +\sum_{\bf i} \Delta_{U}({\bf i}) (f_{{\bf i}\uparrow}^\dagger f_{{\bf i}\downarrow}^\dagger +  g_{{\bf i}\uparrow}^\dagger g_{{\bf i}\downarrow}^\dagger) + {\rm H.c.},
\end{align}
where we have defined the spatially dependent mean-field superconducting order parameter
%
\begin{align}
\label{eq:orderU}
\Delta_{U}({\bf i}) = -U({\bf i})\frac{\langle f_{{\bf i}\downarrow}  f_{{\bf i}\uparrow}\rangle + \langle g_{{\bf i}\downarrow}  g_{{\bf i}\uparrow}\rangle}{2}.
\end{align}
It is straightforward to show that the standard tight-binding Bogoliubov-de Gennes (BdG) formalism (see e.g. Refs.~\onlinecite{Zhu00,Shirai03,Covaci06} for recent applications) applies even in the case of the 2-atom unit cell found in graphene. Written out explicitly, the mean-field Hamiltonian (\ref{eq:H_mf}) can be diagonalized with the following site-dependent two-dimensional Bogoliubov-Valatin transformation
%
\begin{align}
\label{eq:Bogoliubov1}
\left( \begin{array}{c} f_{i\uparrow} \\ g_{i\uparrow} \end{array}\right) & = \sum_{\nu = 1}^{2N}
\left( \begin{array}{c} u_{i}^\nu \\ y_{i}^\nu \end{array}\right) \gamma_{\nu\uparrow} -  
\left( \begin{array}{c} v_{i}^{\nu\ast} \\ z_{i}^{\nu\ast} \end{array}\right) \gamma_{\nu\downarrow}^\dagger \\
\label{eq:Bogoliubov2}
\left( \begin{array}{c} f_{i\downarrow}^\dagger \\ g_{i\downarrow}^\dagger \end{array}\right) & = \sum_{\nu = 1}^{2N}
\left( \begin{array}{c} v_{i}^\nu \\ z_{i}^\nu \end{array}\right) \gamma_{\nu\uparrow} +  
\left( \begin{array}{c} u_{i}^{\nu\ast} \\ y_{i}^{\nu\ast} \end{array}\right) \gamma_{\nu\downarrow}^\dagger,
\end{align}
where $N$ is the number of unit cells in the whole junction, and the resulting diagonal Hamiltonian reads
%
\begin{align}
\label{eq:H_diag}
H_{\rm MF} = \sum_{\nu = 1}^{2N} \sum_\sigma E^\nu \gamma^\dagger_{\nu\sigma}\gamma_{\nu\sigma}.
\end{align}
The energy eigenvalues $E^\nu$ and the eigenfunctions $(u,y,v,z)_{\bf i}^\nu$ are determined by solving the following $4N \times 4N$ eigenvalue problem for its $2N$ positive eigenvalues
%
\begin{align}
\label{eq:eigproblem}
\sum_{\bf j} \left( \begin{array}{cc} H_0({\bf i,j}) & \Delta({\bf i,j}) \\ \Delta^\dagger({\bf i,j}) & -H_0({\bf i,j}) \end{array}\right)
\left( \begin{array}{c} 
u_{\bf j}^\nu \\ y_{\bf j}^\nu \\  v_{\bf j}^\nu \\ z_{\bf j}^\nu
\end{array}\right) 
= 
E^\nu 
\left( \begin{array}{c}
u_{\bf i}^\nu \\ y_{\bf i}^\nu \\ v_{\bf i}^\nu \\ z_{\bf i}^\nu 
\end{array}\right),
\end{align}
%
%
where $H_0$ and $\Delta$ are $2 \times 2$ matrices which for $({\bf i,j}) = ((n,m),(p,r))$ can be written as
%
\begin{align}
\label{eq:H0}
H_0({\bf i,j}) & = \! \left( \begin{array}{cc}  \! \!\!
\tilde{\mu}({\bf i})\delta_{{\bf ij}} & \! \!\!\!\! -t(\delta_{\bf ij} + \delta_{p,n-1}\delta_{r,m\pm1} ) \!\! \\
\! \!\! -t(\delta_{\bf ij} + \delta_{p,n+1}\delta_{r,m\mp1} ) & \tilde{\mu}({\bf i})\delta_{\bf ij} \! \!\!
\end{array}\right) \\
\label{eq:Delta}
\Delta({\bf i,j}) & =  \left(  \begin{array}{cc}  \Delta_{U}({\bf i})\delta_{\bf ij} & 0 \\
0 &  \Delta_{U}({\bf i})\delta_{\bf ij}
\end{array}\right).
\end{align}
As in the standard 1-atom per unit cell BdG formalism, half of the eigenvalues are guaranteed to be positive because $H_0^\ast = H_0$ and $\Delta = \Delta^T$.
The self-consistency condition for the order parameter can finally be written as
%
\begin{align}
\label{eq:orderU2}
\Delta_{U}({\bf i}) = -\frac{U({\bf i})}{2} \sum_{\nu =1}^{2N} (u_{\bf i}^\nu v_{\bf i}^{\nu\ast} +  y_{\bf i}^\nu z_{\bf i}^{\nu\ast})\tanh \frac{\beta E^\nu}{2}.
\end{align}

To significantly reduce the size of eigenvalue problem, we will assume a smooth interface between the S and N regions and use the translational symmetry that exists perpendicular to the junction to apply Bloch's theorem. The formalism and subsequent numerical results are for the zigzag interface (see Fig.~\ref{fig:graphene}) but we have also studied the armchair interface and found no differences. With only on-site superconducting pairing the direction of the interface should indeed not matter.
More specifically, for the zigzag interface we can now write the eigenfunctions as
%
\begin{align}
\label{eq:zzFourier}
\left( \begin{array}{c} u_{{\bf i}=(n,m)} \\ y_{{\bf i}=(n,m)} \\ v_{{\bf i}=(n,m)} \\ z_{{\bf i}=(n,m)}
\end{array}\right) = \frac{1}{\sqrt{N_y}} \sum_{k_y}
\left( \begin{array}{c} u_{n}(k_y) \\ y_{n}(k_y)\\ v_{n}(k_y) \\ z_{n}(k_y)
\end{array}\right)  e^{i k_y m\frac{a}{2}}
\end{align}
where the wave vector $k_y = \frac{2\pi l}{N_y a}$, with $l$ being an integer such that $k_y \in ]\frac{-\pi}{a},\frac{\pi}{a}]$, is a good quantum number for the system and $N_y$ denotes the number of unit cells, of width $T = a = 2.46$~\AA, perpendicular to the interface.
The BdG eigenvalue problem Eq.~(\ref{eq:eigproblem}) now reduces to only depend on the $x$ coordinate indices $(n,p)$ but has to be solved for all $k_y$. 
%

The BdG eigenvalue equations above, Eq.~(\ref{eq:eigproblem}-\ref{eq:Delta},\ref{eq:orderU2}) are solved by starting with an initial guess for the order parameter profile $\Delta_U$ throughout the junction. After finding the $2N_x$ eigenstates of Eq.~(\ref{eq:eigproblem}) for each $k_y$ we can compute a new order parameter profile from Eq.~(\ref{eq:orderU2}). The process is repeated until the difference in order parameters between two subsequent steps are less than a desired accuracy.
The final converged result will allow us to study the proximity effect in graphene. While $\Delta_U$ will always be zero outside S, since per definition it is proportional to $U$, the pairing amplitude
 %
\begin{align}
\label{eq:pairamplitudeU}
F_{U}({\bf i}) & = \frac{\langle f_{{\bf i \downarrow}} f_{{\bf i}\uparrow}\rangle + \langle  g_{{\bf i \downarrow}} g_{{\bf i}\uparrow}\rangle}{2} =-\frac{\Delta_U({\bf i})}{U({\bf i})}
\end{align}
will display the leakage of Cooper pairs into the normal region.

In the above formalism it is the leakage pairing amplitude that is responsible for the Josephson current when the order parameter has a finite phase gradient over the junction. Numerically, we impose a phase gradient by fixing the phase of $\Delta_U$ in the outermost parts of the S regions, labeled $X$ in Fig.~\ref{fig:SNSjunction}, and then solve self-consistently for both phase and amplitude in the remaining SNS junction but only for the amplitude in the $X$ regions. For small currents one should be able to fix the phase in the whole S regions, but physically, as soon as the current is non-zero, there will necessarily also be a phase drop even over the superconducting contacts and not just in the normal region. By extensive testing we have found this to be a non-insignificant effect in many junctions and we have therefore ensured that the region of self-consistency for the phase in the contacts is large enough to ensure bulk-like conditions in S. In terms of calculating the Josephson current vs.~phase drop, $I(\phi)$, we still define the phase variable $\phi$ as the phase drop over the normal region N as is usually done. 

Once a self-consistent solution with a phase drop $\phi$ over the junction is obtained, we calculate the Josephson current  using the continuity equation for the particle current
%
\begin{align}
\label{eq:conteq}
{\bf \nabla \cdot J} + \frac{\partial \rho}{\partial t} = 0
\end{align}
together with the Heisenberg equation
%
\begin{align}
\label{eq:Heisenbergeq}
\frac{{\rm d} n_{\bf i}}{{\rm d}t} = \frac{i}{\hbar}[H,n_{\bf i}],
\end{align}
where $n_{\bf i}$ is the particle density per unit cell with the average $\rho = \langle n \rangle$.
This approach was used in e.g.~Ref.~\onlinecite{Covaci06} for a square lattice. The quantum average of the commutator in Eq.~(\ref{eq:Heisenbergeq}) can easily be shown to only contain the kinetic hopping terms when the self-consistent order parameter is used in the mean-field Hamiltonian (\ref{eq:H_mf}), which is true in our model except in the $X$ regions. The $X$ regions will therefore act as sinks and/or sources for the current. More specifically, we get the Josephson current per cross-sectional distance as 
%
\begin{align}
\label{eq:Icurr}
I(n) = e a_{\rm cell}\tilde{J}(n)/(\hbar a_{\rm xs})
\end{align}
where
%
\begin{align}
\label{eq:Jtilde_zzFT}
\tilde{J}(n) = & -  \frac{8t}{N_y} \sum_{\nu,k_y} \left( {\rm Im}(u_{n}^{\nu\dagger}(k_y)y_{n-1}^\nu(k_y))f(E^\nu) \right. \nonumber \\
& + \left. {\rm Im}(v_{n}^{\nu}(k_y)z_{n-1}^{\nu\dagger}(k_y))f(-E^\nu)\right) \cos(k_y a/2)
\end{align}
and, for the zigzag interface, $a_{\rm cell} = \sqrt{3}a/2$ is the length of the unit cell in the direction of the current  and $a_{\rm xs} = a/2$ is the perpendicular, cross-sectional, distance.

Finally, we will also be interested in the local density of states (LDOS) which easily can be calculated at $T = 0$~K from
 %
\begin{align}
\label{eq:LDOS}
D_{\bf i}(E) = 2\sum_{E^\nu \geq 0} &(|u_{\bf i}^\nu|^2 + |y_{\bf i}^\nu|^2)\delta(E^\nu-E) \nonumber \\
+ &(|v_{\bf i}^\nu|^2 + |z_{\bf i}^\nu|^2)\delta(E+E^\nu).
\end{align}
%
%
%
%
\section{Numerical results}
%
\subsection{Simulation details}
To investigate the proximity effect and Josephson current in graphene SNS junctions we have solved the tight-binding BdG formalism described above self-consistently. The physical input parameters are the on-site pairing potential $U$ in S, the effective potential $\tilde{\mu}$ in S and N, temperature, and the length $L$ of N. More specifically we consider the following set up. For the superconducting contacts: $U(S) = 3.4$~eV $= 1.36t$, $\mu(S) = 1.5$~eV = 0.6$t$. This leads to $\Delta_{U} = \Delta_0 = 0.1$~eV and a superconducting coherence length  $\xi \approx 50$~\AA~$\approx 25$~unit cells in the zigzag direction for the bulk. These values satisfy $\lambda_F(S) \ll \xi$ and allow us to numerically investigate both the $L<\xi$ and $L>\xi$ cases. They are however are quite large values for a realistic situation but smaller superconducting gaps leads to slower convergence rates and also the need for larger systems making calculations less feasible. We have checked our key results for smaller $U$ and found no significant difference. 
For the normal region we have mainly studied short junctions with $L = 10$ unit cells and long junctions with $L = 50$ unit cells with various doping levels. The doping levels have been implemented by setting $\mu(N)$ to values ranging from 0~eV, modeling clean, undoped, graphene where the Fermi level is located at the Dirac point to heavily doped junctions with 1.5~eV which gives no FLM at the interfaces. 
The temperature was chosen to be $T = 10$~K throughout the work, which in comparison with $T_c$ is effectively 0~K.

The accuracy of the solution is determined by the choice of termination criterion for the self-consistency step, the number of $k_y$ points, and the size of S and $X$ to ensure bulk-like superconducting conditions, all of which has been tested thoroughly.
%
%
\subsection{Proximity effect}
\begin{figure}[htb]
\includegraphics[scale = 1]{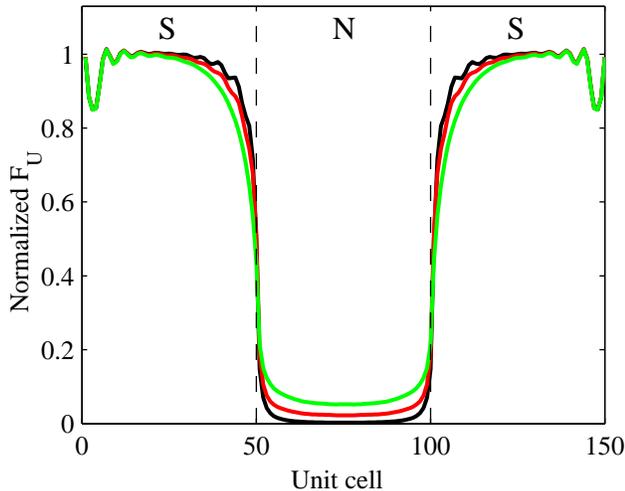}
\caption{\label{fig:proximity} (Color online) Normalized pair amplitude $F_U$ profiles for a S = 50, N = 50 unit cells junction with three different doping levels in N: at Dirac point 0 eV (black), moderate doping 0.7 eV (red), no FLM 1.5 eV (green).}
\end{figure}

Fig.~\ref{fig:proximity} shows typical, normalized, pair amplitude $F_U$ profiles through a SNS junction with different doping levels in N. The depletion of superconducting pairs in S close to the junction as well as the leakage of pairs into the junction is clearly visible. The oscillations at each end of S are due to the end surfaces of the contacts, and are not of primarily importance here. For small doping levels, there are also pronounced oscillations in $F_U$ at the interface on the S side. These oscillations are correlated with oscillations in the charge density which are present even for the case $U = 0$ and therefore attributed to Friedel-like charge oscillations due to the FLM at the interface.
%
%
%
%
\subsection{I vs.~${\bf \phi}$ curves}
As Fig.~\ref{fig:proximity} clearly shows there exists a pronounced proximity effect in graphene SNS junctions. This effect is obviously not taken into account in a calculation where the order parameter is assumed to change abruptly at the interface, such as in Ref.~\onlinecite{Titov06}, and the natural question arises if this effect will significantly change any predictions made by such a non-self-consistent calculation. Among the most significant quantities of a SNS junction is the Josephson supercurrent it can sustain if the junction is short enough to allow coherent transport of superconducting pairs. We investigate in the following two subsections two properties of the Josephson current. In this subsection we extract the current vs.~phase relationship $I(\phi)$ for junctions with different lengths and doping levels and in the next subsection we will look at the length dependence of the critical current for short junctions. Both of these results were extensively worked out in the DBdG non-self-consistent formalism in Ref.~\onlinecite{Titov06}.
 
Before proceeding it should be noted that the largest phase drop we can put over the whole structure is $\pi$. However, with a non-zero phase drop over the S regions the maximum phase drop $\phi$ over the junction will be smaller than $\pi$ for junctions with a significant current and we will in these cases not be able to numerically trace out the $I(\phi)$ relationship over the full $[0,\pi]$ interval. While this appears as a numerical artifact in this context it is in fact closely related to the physical $2\pi$ phase-slip process in Josephson junctions (see e.g.~Ref.~\onlinecite{Tinkhambook}).

\begin{figure}[htb]
\includegraphics[scale = 1]{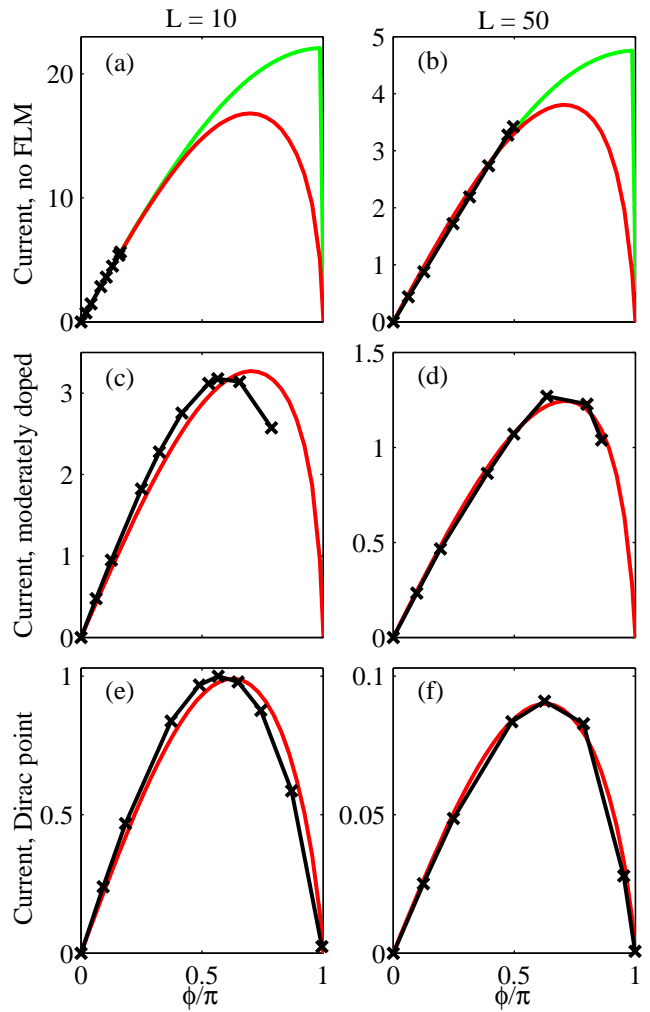}
\caption{\label{fig:Ivsphi} (Color online) $I(\phi)$ relationsship, normalized with respect to the critical current in (e) for short $L<\xi$ (a,c,e) and long $L>\xi$ (b,d,f) junctions when the N region is doped at the Dirac point $\tilde{\mu} = 0$~eV (e,f), moderately doped $\tilde{\mu} = 0.7$~eV (c,d), and with no FLM $\tilde{\mu}=1.5$~eV (a,b). Self-consistent numerical data are in black with the lines being guides to the eye. Red curves are least squares fits (of the overall constant) to the DBdG results whereas green curves are fits to the functional form ${\rm sgn}(\cos(\phi/2))\sin(\phi/2)$.}
\end{figure}

Fig.~\ref{fig:Ivsphi} shows the phase dependence of the Josephson current for both short junctions ($L <\xi$) and long junctions ($L > \xi$) for three different doping levels; undoped $\tilde{\mu}(N) = 0$~eV, moderately doped such that $\tilde{\mu}(N) = 0.7$~eV~$ \gg \hbar v_F/L$, and with no FLM at the interface. The currents have been normalized to the maximum value of the undoped, short junction. 
%
The self-consistent data are in black whereas the red and green curves are least squares fits to results obtained in the DBdG formalism. 
More specifically, for the undoped junction the red curve is a fit, with $C$ as the fitting parameter, to Eq.~(20) in Ref.~\onlinecite{Titov06}:
 %
\begin{align}
\label{eq:Titovmu0}
I(\phi) = C \cos(\phi/2){\rm arctanh}[\sin(\phi/2)].
\end{align}
This phase relationship is notably different from the traditional Josephson form\cite{Josephson64}, $I = I_c \sin(\phi)$, but is in fact identical to that of a disordered metal upon substitution $k_Fl \rightarrow 1$, where $l$ is the mean free path. This is despite the fact that the graphene SNS junction is treated as a ballistic junction and is here instead a consequence of the Dirac band spectrum of graphene. As can be seen in Fig.~\ref{fig:Ivsphi} there is good agreement to this functional dependence for the short junction at the Dirac point (e), but also for the long junction (f) where the approximation of calculating the Josephson current from the subgap Andreev bound states in the DBdG formalism is not formally motivated.
For non-zero, but moderate, doping of the junction a closed form analytical result from the DBdG formalism is not available, though the functional dependence on $\phi$ is still close to Eq.~\ref{eq:Titovmu0}, and there is still good agreement with our self-consistent results (Fig.~\ref{fig:Ivsphi} c,d).
Finally, for the case of no FLM between the S and N regions (Fig.~\ref{fig:Ivsphi} a,b), the green curves are fit to the functional form ${\rm sgn}(\cos(\phi/2))\sin(\phi/2)$ which was derived recently by Linder {\it et.~al}\cite{Linder07com}. Due to the high currents in the no FLM cases, it is not possible to obtain current values for phase drops larger than those reported in Fig.~\ref{fig:Ivsphi}, and it is thus impossible to distinguish which fits are best. 
In fact, for both of these two junctions as well as for the longest junctions we could conveniently model ($L \approx 210$~\AA) with no FLM, the superconductivity energy gap is never depleted inside the junction, and, subsequently, the phase drop profile across the whole structure is linear. A linear phase drop across the whole structure will give a linear relationship between $I$ and $\phi$ which is clearly seen in Fig.~\ref{fig:Ivsphi}. This persistent superconductivity gap in the junction has the intriguing consequence that for these junctions the normal region has, at least in our structure sizes, a diverging Ginzburg-Landau coherence length, despite the fact that the pairing potential is zero inside the junction. 
%
%
\subsection{$L$ dependence of the critical current}
In the previous subsection the I vs.~$\phi$ dependence was investigated and fitted to analytical forms derived from the non-self-consistent DBdG formalism. No attention was however paid to the dependence of the prefactor, i.e.~the fitting parameter, or equivalently the critical current, on various physical quantities such as doping level in N, junction length, gap size etc. In this subsection we will focus on the behavior of the critical current $I_c$ on the length of the junction $L$ in the short junction regime. More specifically, we will investigate the influence of the proximity effect on the $L$ dependence of the following two results derived using the DBdG formalism\cite{Titov06}
 %
\begin{align}
\label{eq:TitovLdep0}
I_c & = 1.33\frac{e\Delta}{\hbar}\frac{1}{\pi L} \ \ \ {\rm when} \ \tilde{\mu} = 0,\\
\label{eq:TitovLdep97}
I_c & = 1.22\frac{e\Delta}{\hbar}\frac{\mu}{\pi \hbar v_F} \ \ \ {\rm when} \ \tilde{\mu}\gg\hbar v_F/L.
\end{align}
%
%
\begin{figure}[htb]
\includegraphics[scale = 1]{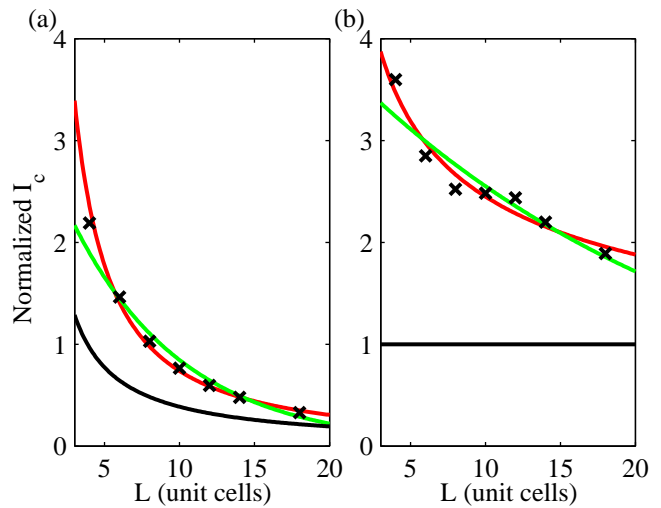}
\caption{\label{fig:Ldep} (Color online) Normalized critical Josephson current $I_c$ as function of length $L$ of the junction at (a) the Dirac point $\tilde{\mu} = 0$~eV and (b) moderately heavy doping $\tilde{\mu}=0.7$~eV. Self-consistent results (crosses), Eqs.~(\ref{eq:TitovLdep0}-\ref{eq:TitovLdep97}) (black), and least-square fits to the functional forms $CL^{b}$ (red) and $Ce^{-L/\xi_N}$ (green). The current is normalized such that the critical current in Eqs.~(\ref{eq:TitovLdep97}) is equal to 1.
}
\end{figure}
Fig.~\ref{fig:Ldep} shows the the $I_c$ vs.~$L$ dependence for the self-consistent results (crosses) and Eqs.~(\ref{eq:TitovLdep0}-\ref{eq:TitovLdep97}) (black), together with least squares fits to the functional forms $CL^{b}$ (red) and $Ce^{-L/\xi_N}$ (green). The current is normalized such that $I_c(L=10) = 1$ for Eq.~(\ref{eq:TitovLdep97}).
For a junction at the Dirac point, Fig.~\ref{fig:Ldep}(a), we see that the proximity effect causes an increase in the critical current as well as change the L dependence from $b=-1$ in Eq.~(\ref{eq:TitovLdep0}) to $b = -1.3$. Functionally this form is also close to the traditional Ginzburg-Landau $Ce^{-L/\xi_N}$ functional dependence with $\xi_N = 16$~\AA. 
At moderate doping, here represented by $\tilde{\mu} = 0.7$~eV $\gg\hbar v_F/L$, Fig.~\ref{fig:Ldep}(b) shows that the effect of a self-consistent calculation is even more pronounced. Eq.~(\ref{eq:TitovLdep97}) has no $L$ dependence whereas we see a clear increase in the critical current when the junction size decreases. The fits are not as good as in the undoped case but $b \approx -0.4$ or $\xi_N \approx 50$~\AA. Also note the significant increase, between 2 and 4 times, in the current for all junction lengths investigated.
Interestingly, the difference in $b$-exponents between the two cases are $\sim 1$, just as the DBdG prediction, although both values are significantly modified in our self-consistent calculations.

%
\subsection{LDOS for SN junction}
\begin{figure}[htb]
\includegraphics[scale = 1]{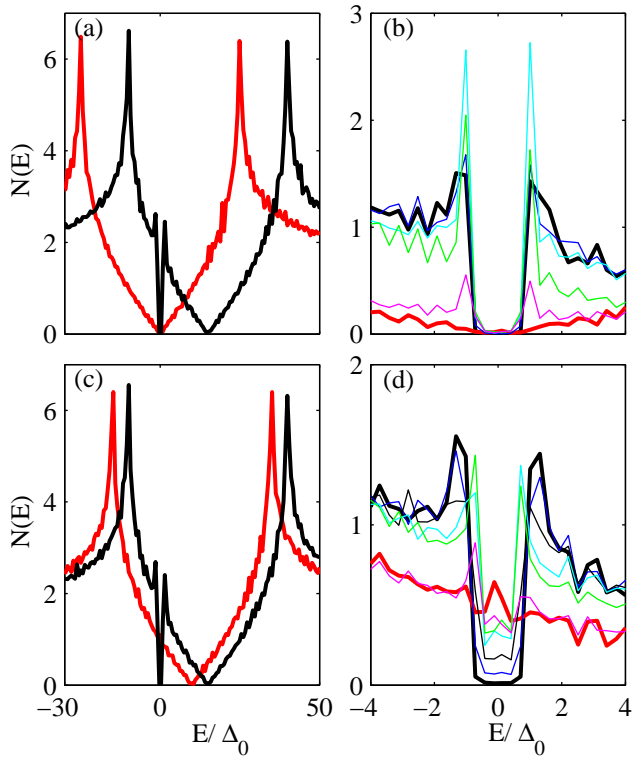}
\caption{\label{fig:LDOS} (Color online) Typical LDOS for SN junction as function of the normalized energy defined with reference to the Dirac point. Upper plots (a,b):  N region is undoped, at the Dirac point. Lower plots (c,d): N is moderately doped. Right hand plots are magnifications of the gap features of the left hand plots. Different colors represent different positions inside the junction: middle of S (black), middle of N (red), interface (green), interface $\pm 2$ unit cells (magenta, cyan), interface $- 20$ unit cells (blue).}
\end{figure}

Finally, we report in Fig.~{\ref{fig:LDOS}} on how the local density of states (LDOS) evolve through a graphene SN interface with an abrupt effective chemical potential $\tilde{\mu}$ change at the interface. 
Since for a non-uniform system, such as a SN junction, the superconducting energy gap will in general not be equivalent to the order parameter $\Delta_U$ in the full self-consistent solution we are especially interested in the evolution of the superconducting energy gap and the adjoining coherence peaks.
The relevant superconducting length scale is the coherence length which is $\xi \approx 50$~\AA~$\approx 25$~unit cells on the S side. 
%
The two upper plots (a,b) show the LDOS when the normal region is undoped, with the right hand plot being a magnification around zero energy, defined to be at the Dirac point. As can be seen, the superconducting coherence peaks on each side of the energy gap are pronounced, and in fact enhanced close to the interface, all the way up to the interface but die out very quickly, over only a few unit cells, on the N side. Since the DOS is zero at the Dirac point it is hard with the current resolution to determine the exact evolution and completeness of the superconducting energy gap.
For the two lower plots (c,d) the N region is instead biased into a moderately heavy doping regime. Here the superconducting gap is not complete even at distances $\sim 40$~\AA~from the interface on the S side. Close to the interface the coherence peaks are shifted towards lower energies. On the N side of the interface a normal state LDOS is again achieved within only a few unite cells. 

The above results thus give a very short decay length for the superconducting state in the N region but show on large and spatially extended effects of the superconducting gap on the superconducting S side, especially as we move away from the Dirac point.
This should be contrasted with the results for junctions with no FLM where a superconducting state, with an complete energy gap, persist in N even in the longest junctions we can model.
The significant different response between undoped or moderately doped junctions on one hand and junctions with no FLM on the other hand should be experimentally accessible using a point contact scanning probe to investigate the LDOS. 

%
%
\section{Conclusions}
Using the TB BdG formalism we have been able to calculate the proximity effect and Josephson current in ballistic SNS graphene Josephson junctions. 
We have shown that the functional form of $I(\phi)$ derived using the non-self-consistent DBdG formalism for short junctions\cite{Titov06} is qualitatively valid at both in undoped and in moderately doped junctions. In addition, we have demonstrated that these results are extendable to the long junction regime where the junction length $L$ is longer than the superconducting coherence length.
However, the dependence on the junction length $L$ for the critical current in short junctions is enhanced, from $L^{-1}$ to $L^{-1.3}$ for undoped and from $L^0$ to $L^{-0.4}$ for moderately doped junctions, when proximity effect is taken into account. Also, the magnitude of the critical current is enhanced.
For junctions with no FLM between S and N, i.e.~the same doping level throughout the structure, superconductivity is not depleted inside N even for the longest junctions we were able to investigate. This means that in a ballistic, no FLM, graphene SNS junction the Ginzburg-Landau superconducting coherence length in the normal region is close to diverging. This intriguingly indicates that the normal region is close to a superconducting instability, despite the fact that no pairing potential is present in this region.
This behavior is in sharp contrast with the LDOS data for SN junctions when N is undoped or only moderately doped where the superconducting gap disappears within only a few unit cells on the N side of the junction.
%
%
%
\begin{acknowledgments}
We thank Mac Beasley and Jacob Linder for valuable discussions.
A.M.B.-S. acknowledges partial support from DOE contract DE-AC02-76SF00515.
\end{acknowledgments}


\end{document}